 \documentstyle[prl,twocolumn,floats,aps]{revtex}

\tighten
\begin{document}
\input epsf  
\preprint{Draft \today}
\draft
\bibliographystyle{prsty}  

%
%
%
%
%
%

\title{Time Delay in the Kuramoto Model of Coupled Oscillators}
\author{M. K. Stephen Yeung and Steven H. Strogatz}
\address{Department of Theoretical and Applied Mechanics, \\
	 Kimball Hall, Cornell University, Ithaca, NY 14853-1502}
\date{\today}
\maketitle

\begin{abstract}
We generalize the Kuramoto model of coupled oscillators to allow
time-delayed interactions.
New phenomena include bistability between synchronized and
incoherent states, and unsteady solutions with time-dependent order
parameters.
We derive exact formulas for the stability boundaries of the 
incoherent and synchronized states, as a function of the delay, in 
the special case where the oscillators are identical.  
The experimental implications of the model are discussed for 
populations of chirping crickets, where the finite speed of sound 
causes communication delays, and for physical systems such as coupled 
phase-locked loops or lasers.
\end{abstract}


\narrowtext



The Kuramoto model of coupled oscillators is one of the most celebrated
systems in nonlinear dynamics.
It was originally developed as an analytically tractable version of
Winfree's mean-field model for large populations of biological 
oscillators \cite{Winfree67}, such as groups of chorusing crickets 
\cite{cricket_ref}, flashing
fireflies \cite{Buck88}, and cardiac pacemaker cells \cite{Peskin75}.
In a beautiful analysis, Kuramoto showed that the model exhibits a
spontaneous transition from incoherence to collective synchronization, 
as the coupling strength is increased past a certain threshold
\cite{Kuramoto75_84}.
The model has since been analyzed more deeply and extended in various 
ways 
\cite{SK86_S88,SM_Crawford,SMM92,BPS98,kura_ref}.
It has also been linked to several physical problems,
including Landau damping in plasmas \cite{SMM92}, the dynamics of
Josephson junction arrays \cite{WCS96_98}, bubbly fluids
\cite{Smereka_ref}, and coupled Brownian ratchets \cite{HBH97}.

Here we explore the effects of time delay on the dynamics of 
the Kuramoto model.  In the past, delay has often been
neglected in models of 
coupled oscillators.  In many cases this approximation is physically 
justified, and in all cases it simplifies the mathematics.   
But recently several authors
have begun to investigate oscillator systems where delays are not
negligible \cite{NSK91,delay_ref}, 
motivated by neural networks where synaptic, dendritic, 
and propagation delays can be significant.  Other authors 
have considered delays in systems of limit-cycle oscillators 
\cite{limit_cycle_with_delay_ref},
with applications to 
arrays of lasers and microwave oscillators.

Intuitively, the problem is similar to that faced by the fans sitting 
in an enormous football stadium, all of whom (we suppose) are trying 
to clap in unison.  Even if everyone were successfully clapping in 
perfect synchrony, it would not sound that way to the fans 
themselves, as the applause coming from far across the field would be 
significantly delayed, because of the finite speed of sound.

Nevertheless, we show that perfect synchrony is possible in the 
Kuramoto model with time delay, if all oscillators are identical.  
In fact, there can be 
several different synchronized states, and they can co-exist with 
a stable incoherent state where the oscillators are completely 
disorganized.  These multistabilities are qualitatively new: 
they do not occur in the original Kuramoto model.


We consider a system of phase oscillators with noisy, randomly 
distributed intrinsic frequencies, and with delayed mean-field 
coupling:
\begin{equation}
    \dot{\theta_i} (t)
  = \omega_i + \xi_i (t) + {K \over N} \sum_{j=1}^N
      \sin ( \theta_j (t-\tau) - \theta_i (t) - \alpha ),
\label{model}
\end{equation}
for $i=1,\dots,N.$
Here $\theta_i (t)$ is the phase of the $i$th
oscillator at time $t$, and $\omega_i$ is its intrinsic frequency, 
randomly drawn from a probability
density $g(\omega)$ with mean $\omega_0.$ The white noise $\xi_i (t)$
represents frequency fluctuations at an effective temperature 
$D \geq 0,$ and is defined by the
ensemble averages $< \xi_i(t) > = 0,$
$< \xi_i(s) \xi_j(t) > = 2 D \delta_{ij} \delta(s-t).$
In the global coupling term, $K \geq 0$ is the coupling strength,
$\tau > 0$ is the delay, and $\alpha$ is a phase frustration parameter.
This model reduces to the Kuramoto model
\cite{Kuramoto75_84}
if $\tau=0$, $\alpha=0$, and $D=0$, and to the mean-field XY
model if $\tau=0$, $\alpha=0$, and the oscillators are identical, i.e.,
$g(\omega) = \delta(\omega - \omega_0).$  For $\tau=0$, the separate 
effects of frustration $\alpha$ and noise $D$
have been studied by Sakaguchi and Kuramoto
\cite{SK86_S88}.

As the one-parameter 
family of rotating-frame transformations
$ \theta_i (t) \rightarrow \theta_i (t) - \Omega t,
  \omega_i \rightarrow \omega_i - \Omega,
  \alpha \rightarrow \alpha + \Omega \tau $
leave Eq.\ (\ref{model}) invariant for any $\Omega,$ we may 
assume $\alpha = 0$ without loss of generality --- except if 
$\tau = 0$, which we forbid.  (This restriction is merely for 
convenience. 
All our results are well-behaved as $\tau \rightarrow 0$ and 
converge to those obtained by setting $\tau = 0$ from the start.)  
Moreover, since Eq.\ (\ref{model}) is invariant under the 
reflection
$ \omega_i \rightarrow - \omega_i, $
$ \theta_i \rightarrow - \theta_i, $
$ \alpha \rightarrow - \alpha, $
it suffices to consider $\omega_0 \geq 0$.

It is often helpful to describe the macroscopic state of the system in
terms of the complex order parameter
$R(t) e^{i \psi(t)} = {1 \over N} \sum_{j=1}^N e^{i \theta_j (t)}$
introduced by Kuramoto
\cite{Kuramoto75_84}.
Here $R(t)$ measures the system's phase coherence.  In particular, 
$R=1$ if all the oscillators are in phase, whereas $R = 0$ if 
the oscillators are scattered around the
unit circle with their centroid at the origin.


Our first analytical result concerns the stability of the incoherent 
state for the infinite-$N$ limit of Eq.\ (\ref{model}).  We rewrite 
the model as a Fokker-Planck equation for the density 
$\rho(\theta, \omega, t)$ of oscillators currently at phase $\theta$, 
with intrinsic frequency $\omega$.
Because the method is standard \cite{SM_Crawford,SMM92},
we omit the details.  The only new feature here is that the drift 
velocity field inherits the time delay in Eq.\ (\ref{model}).  When the
Fokker-Planck equation is linearized about the incoherent state (the
stationary density $\rho \equiv 1/2 \pi$), we find \cite{tobe} that 
its continuous spectrum is
$  \{ -D - i \omega \ | \ \omega \in {\mathrm supp}(g) \}. $
Hence for $D>0,$ the continuous spectrum corresponds to damped modes 
and therefore the stability of the incoherent state is
determined solely by the discrete eigenvalues.  But when $D=0,$ the
continuous spectrum is pure imaginary and corresponds to
neutrally stable rotating waves in the full system.  In this case, the
incoherent state can never be linearly stable: it is either unstable 
or neutral, depending on the discrete eigenvalues.  These eigenvalues
$\lambda$ satisfy \cite{tobe} 
\begin{equation}
  e^{-\lambda \tau} {K \over 2} \int_{-\infty}^{\infty} d \omega
     {g(\omega) \over \lambda + D + i \omega} = 1.
\label{eigen_eqn}
\end{equation}

This implicit formula for $\lambda$ is exact but difficult to analyze 
for arbitrary $g(\omega)$, so we consider the case of identical 
oscillators to gain some insight.
Even this case turns out to be far from trivial.
If $g(\omega) = \delta(\omega - \omega_0),$ Eq.\ (\ref{eigen_eqn})
can be simplified to the transcendental equation
\begin{equation}
  (p + i r - z) e^z + q = 0,
\label{exp_polynomial}
\end{equation}
where
$p = -D \tau \leq 0, r = - \omega_0 \tau, q = K \tau /2,$ and
$z = \lambda \tau.$
Then the stability of the incoherent state depends on whether all 
roots of Eq.\ (\ref{exp_polynomial}) satisfy ${\mathrm Re}(z)<0$ 
(in which case 
we will say ``all eigenvalues are stable'', for brevity).

By the transformations
$  z \rightarrow z + i n \pi,
   q \rightarrow (-1)^n q,
   r \rightarrow r + n \pi, $
and
$  z \rightarrow z^*,
   r \rightarrow -r, $
we may assume $r \in [0,\pi/2]$ in Eq.\ (\ref{exp_polynomial}).
For $r=0,$ Hayes proved \cite{Hayes50,BC63} that
all eigenvalues are stable
if and only if
$ p < 1 $ and $ p < -q < \sqrt{p^2 + {y_1}^2}, $
where $y_1$ is the unique zero of $p \sin y - y \cos y$ in $(0,\pi).$
Using results of Pontryagin as in Ref.\ \cite{BC63}, we can 
show \cite{tobe,p<=0} that for $r \in (0,\pi/2],$
\begin{itemize}
  \item  if $p=0,$ then all eigenvalues are stable
             if and only if $r - {\pi / 2} < q < 0.$
  \item  if $p<0,$  then all eigenvalues are stable if and only if
	 $-\sqrt{p^2 + (y_2 - r)^2} < q < \sqrt{p^2 + (y_1 - r)^2},$
	 where $y_1$ and $y_2$ are the unique zeroes of
	 $p \sin y + (r-y) \cos y$ in $(0,r)$ and $(\pi/2,\pi)$
	 respectively.
\end{itemize}
These conditions are exact but still opaque, so we simplify the model 
further for illustration.
Suppose there is no noise ($D=0$).  Then we find \cite{tobe} that the
incoherent state is neutrally stable precisely when
\begin{equation}
    K < {\omega_0 \over 2m-1} \ \ {\mathrm and} \ \
    {(4 m - 3) \pi \over 2 \omega_0 - K} < \tau <
    {(4 m - 1) \pi \over 2 \omega_0 + K},
\label{inco_stable}
\end{equation}
with $m$ being an arbitrary positive integer.

\begin{figure}[hbt]
  \centerline{\epsffile{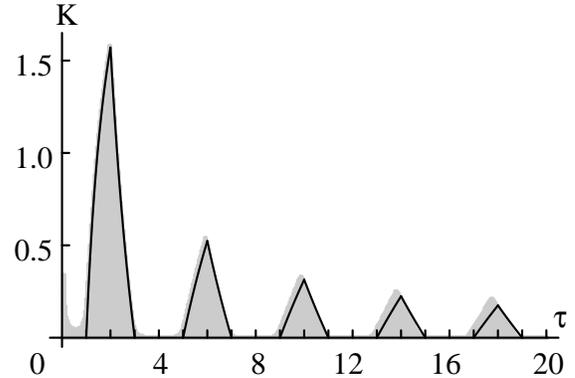}}
\caption{Stability region for the incoherent state, with
	   $g(\omega) = \delta(\omega - \omega_0),$
	   $\omega_0 = \pi/2, D=0, N=12.$
	   Black curves: theoretical boundaries (\ref{inco_stable}) for 
	   the infinite-$N$ limit;
	   Grey area: results from numerical integration using a sixth
           order Adams-Bashforth-Moulton scheme, with fixed stepsize 
	   $dt=\tau/20$, and with the corrector formula iterated for 
	   convergence and stability.  Initially, all the phases were 
	   evenly spaced, and the symmetry was broken by adding 
	   $O(10^{-10})$ random perturbations.  The incoherent state 
	   was judged as unstable if $R(t) > 10^{-7}$ at a final time 
	   of $t=800 \tau$.}
\label{fig_inco}
\end{figure}

Figure \ref{fig_inco} shows that this analytical result agrees with
numerical simulations, even for as few as $N=12$ oscillators.  
Although the incoherent state is neutrally stable in the grey 
region, we observe numerically that 
$R(t) \rightarrow 0$ exponentially fast, as in Landau damping 
\cite{SMM92}.  In this sense,
incoherence is stable in the grey region.

Continuing with the instructive case of noiseless, identical 
oscillators, we now consider the possibility of perfect 
synchrony: $\theta_i(t) = \theta (t)$ for all $i.$  We restrict 
our attention to a particular class of such solutions, namely
uniform rotations:
$\theta (t) = \Omega t + \beta.$ 
Self-consistency then requires
\begin{equation}
  \Omega = \omega_0 - K \sin (\Omega \tau)
\label{sync_freq}
\end{equation}
for such solutions to exist, and linearization \cite{tobe} imposes
\begin{equation}
  \cos (\Omega \tau) > 0
\label{sync_stable}
\end{equation}
as the condition for their orbital stability.

If we graph both sides of Eq.\ (\ref{sync_freq}) as functions of 
$\Omega$, we see that for all sufficiently large $K,$ there exist 
multiple stable synchronized
states \cite{small_tau}, as Eq.\ (\ref{sync_freq}) has non-unique
solutions satisfying (\ref{sync_stable}).
We can also see that stable synchrony is {\em impossible} for certain
combinations of $\tau$ and $K$.  The problem reduces to characterizing 
the two-parameter family of lines of negative slope that intersect the 
sine function on its descending limbs.  We find that stable 
synchronized states do not exist if and only if

\begin{equation}
    K < {\omega_0 \over 2 (2m-1)} \ \ {\mathrm and} \ \
    {(4 m - 3) \pi \over 2 \omega_0 - 2K} < \tau <
    {(4 m - 1) \pi \over 2 \omega_0 + 2K},
\label{sync_unstable}
\end{equation}
with $m$ being an arbitrary positive integer.

\begin{figure}[hbt]
  \centerline{\epsffile{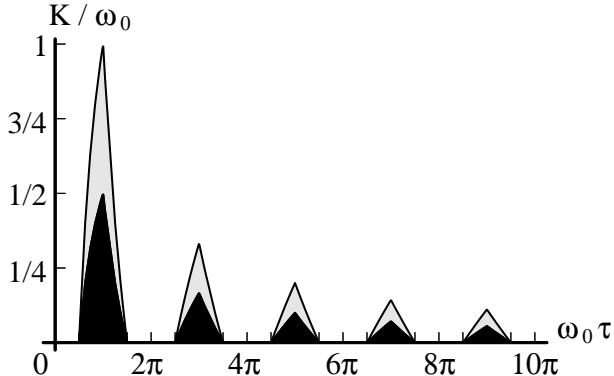}}
  \caption{Stability regions of the incoherent state
           ((\ref{inco_stable})) and the synchronized states
           ((\ref{sync_unstable})) for
           $g(\omega) = \delta(\omega - \omega_0),$ $D=0.$
           White region: one or more stable synchronized states
           exist but the incoherent state is unstable;
           Black region: incoherence is stable but synchrony is not;
           Grey region: one or more stable synchronized states coexist
	   with stable incoherence.}
\label{fig_bistability}
\end{figure}

These zones of forbidden synchrony are shown in black in
Fig.\ \ref{fig_bistability}.  For comparison, they are overlaid on top 
of the earlier grey regions (Fig. \ref{fig_inco}) where incoherence is 
stable.  The black regions fit neatly inside the grey; they have the 
same base and
half the height.  The exposed parts of the grey regions correspond to
bistability: stable incoherence coexists with stable synchrony, and
hysteresis can occur.

\begin{figure}[hbt]
  \centerline{\epsffile{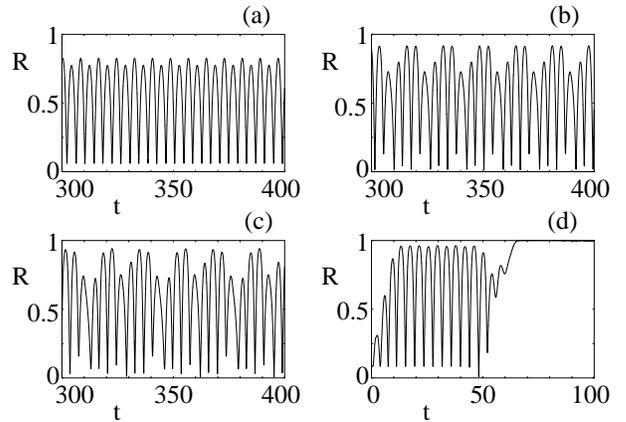}}
  \caption{Time series showing nonsteady order parameter $R(t),$
           with  $g(\omega) = \delta(\omega-\omega_0),$
           $\omega_0 = \pi/2, D=0, \tau=2, N=24.$
           (a) $K = 1.3:$ period-2 oscillation;
           (b) $K = 1.4:$  period-4 oscillation;
           (c) $K = 1.44:$ period-8 oscillation;
           (d) $K = 1.475:$ $R(t) \rightarrow 1$ after 
               a periodic transient.}
\label{fig_nonconstant}
\end{figure}

Numerical simulations reveal windows in the bistable regions where
$R$ can be time-periodic (Fig.\ \ref{fig_nonconstant}). In the 
example shown, period doubling occurs as $K$ increases, but seems to 
be truncated beyond period 16 (not shown); after that, a synchronized 
state apparently takes over, suppressing further
period doubling (Fig.\ \ref{fig_nonconstant}(d)).
Such unsteady behavior is a consequence of the delay; in the standard
Kuramoto model, numerical experiments show that $R(t)$ always
approaches a constant value if $g(\omega)$ is
unimodal and symmetric (although this has never been proven).
Oscillator configurations with two or more clusters
\cite{cluster_ref}
cause the unsteady behavior seen here.  All the clusters move with the 
same average velocity, but their separation is periodically modulated.

\begin{figure}[hbt]
  \centerline{\epsffile{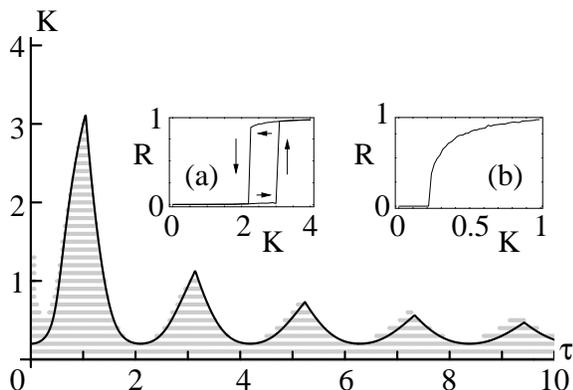}}
  \caption{Stability region of the incoherent state for Lorentzian
           $g(\omega)$ with
	   $\omega_0 = 3,$ $\gamma=0.1,$ $D=0,$ $N=3600.$
           Black curves: theoretical boundary Eq.\
           (\ref{lorentzian_stability});
           Grey strips: numerical results using the same method as in
	   Fig.\ \ref{fig_inco}.
	   Insets:
	   (a) Subcritical Hopf bifurcation of the incoherent state
	       at $\tau=1.$  
	       Arrows indicate a hysteresis loop between
	       stable incoherence and a stable partially locked
	       state with $R$ close to $1.$  
           (b) Supercritical Hopf bifurcation of the incoherent state
	       at $\tau=2.$  A stable partially locked state grows
	       continuously from the incoherent state with 
	       $R = O(\sqrt{K - K_c}).$
	   }
\label{fig_lorentzian}
\end{figure}

So far we have concentrated on identical oscillators.  To check how the
results would be modified for other frequency distributions, we have
considered the Lorentzian distribution
$
g(\omega) = (\gamma / \pi)
	    (\gamma^2 + (\omega - \omega_0)^2)^{-1}.
$
Then by a remarkable coincidence \cite{tobe}, Eq.\ (\ref{eigen_eqn}) 
can again be reduced to
Eq.\ (\ref{exp_polynomial}), but now with
$p = - (\gamma + D) \tau, r = - \omega_0 \tau, q = K \tau /2,$ and
$z = \lambda \tau.$
The critical coupling is given by
\begin{equation}
    K_c = 2 (\gamma + D) \sec (\Omega_c \tau),
\label{lorentzian_stability}
\end{equation}
where
$\Omega_c = \omega_0 - (\gamma + D) \tan (\Omega_c \tau).$
Figure \ref{fig_lorentzian} plots the corresponding region where
incoherence is stable.  It resembles Fig.\ \ref{fig_inco}, with a 
series of evenly spaced peaks (at $\omega_0 \tau = (2n+1) \pi$)
that decrease in height. The main difference is that the distributed 
frequencies produce some rounding of the boundary, and lift it off 
the $\tau$-axis so that it now has minima
$2 (\gamma + D)$ at $\omega_0 \tau = 2n \pi.$

The bistability found earlier also has a counterpart in the Lorentzian 
case (Fig.\ \ref{fig_lorentzian}(a)). 
Partially locked states (which may not be unique)
replace the earlier in-phase states, 
but otherwise the story is unchanged \cite{no_periodic_R}.
Thus, the case of identical oscillators captures the essential features
introduced by delay.

Our final result concerns the bifurcation at $K = K_c$,  where the
incoherent state becomes unstable.  We have adapted the two-timing 
method of Ref. \cite{BPS98} to handle the delay-differential equations
(\ref{model}).   We find \cite{tobe} that generically, for $D \geq 0$
and arbitrary $g(\omega),$ a Hopf bifurcation occurs at 
$K_c$ \cite{special_values}, giving rise to a partially locked
state, or in the density description, a rotating wave with constant
coherence $R = O(\sqrt{|K - K_c|}).$
This bifurcation may be subcritical (Fig.\ \ref{fig_lorentzian}(a)) or
supercritical (Fig.\ \ref{fig_lorentzian}(b)).


Experimental tests of the model may be possible in arrays of 
phase-locked loops \cite{LS78}, relativistic magnetrons
\cite{BSWSH89}, or solid-state lasers \cite{FCRL93},
as they are all approximately governed by coupled Adler 
equations \cite{Adler46} similar in form to Eq.\ (\ref{model}). The 
delay and the coupling strength are both natural control parameters, 
and 
perhaps one could try to map out the stability boundaries, look for 
hysteresis between incoherence and synchrony, etc.  Our model may also
help to explain how crickets can synchronize their chirps 
\cite{cricket_ref}, 
despite the time delays caused by the speed of sound.
Crickets listen to each other's chirps and adjust their own timing
according to a phase response curve \cite{cricket_ref}.
The propagation delay between two crickets 3 meters
apart is about 10 msec.  This is 
short compared to the chirp period (300-500 msec).
Our results suggest that 
delay effects become significant only in the first peak in 
Fig.\ \ref{fig_bistability}, i.e., for delays near half the period of 
the oscillation.  Thus the delays that crickets actually encounter in 
the field
are probably negligible as far as synchrony is concerned.  It would be
interesting to try lab experiments on crickets interacting via chirp
signals whose delay and amplitude can be electronically manipulated.

Research supported in part by the National Science Foundation.
We thank Tim Forrest for information about crickets.

%

%
%


\begin{thebibliography}{10}
\bibitem{no_periodic_R}
However, we have not seen unsteady $R(t)$ in simulations for Lorentzian
$g(\omega)$.
\bibitem{special_values}
At special parameter values, e.g., the turning points 
$\omega_0 \tau = n \pi$ of the stability boundary in 
Fig.\ \ref{fig_lorentzian}, Eq.\ (\ref{model}) is $O(2)$-symmetric and 
codimension-2 bifurcations occur.
\end{thebibliography}
\end{document}